\documentstyle[12pt]{article}
\newtheorem{conjecture}{Conjecture}[section]
\newtheorem{theorem}[conjecture]{Theorem}
\newtheorem{lemma}[conjecture]{Lemma}

\newtheorem{definition}[conjecture]{Relations}
\newtheorem{remark}[conjecture]{Remark}
\newtheorem{example}[conjecture]{Example}

\font\smmm=cmr7
\begin{document}\baselineskip20pt
\title{Roots of Unity: Representations of Quantum Groups}
\author{W. A. Schnizer\thanks{Supported by the Japan Foundation for
the Promotion of Science}\\
RIMS, Kyoto University, Kyoto 606, Japan\\
and\\
Institute for Nuclear Physics, TU-Wien, 1040 Wien, Austria}
\date{February 19, 1993}
\maketitle
\begin{abstract}$\gdef\e{\varepsilon}$$\gdef\s{\sigma}$
$\gdef\ot{\mathop{\otimes}}$
$\gdef\E{{\cal E}}$
$\gdef\F{{\cal F}}$
$\gdef\T{{\cal T}}$
Representations of Quantum Groups $U_{\e}(g_n)$, $g_n$
any semi simple Lie algebra of rank $n$,
are constructed from arbitrary representations
of rank $n-1$ quantum groups for $\e$ a root of unity.
Representations which have the
maximal dimension and number of free parameters for irreducible
representations arise as special cases.
\end{abstract}

\def\<{\langle}
\def\>{\rangle}
\def\sect#1{\advance\count80 by 1 \vskip1cm
\parindent0pt{\large\bf \number\count80.)\, #1}\\\\
\parindent0.4cm}
\sect{{ Introduction}}
Deformations of semi simple Lie algebras \cite{DR,JI} appear as a common
algebraic structure in the field of low dimensional integrable
systems. In many cases the deformation parameter is an $N$-th root
of unity, where $N$ can correspond e.g. to the number of states per site
or to the lattice size in a two dimensional  model.
We will denote the deformation parameter
by $\varepsilon$, if the parameter is an $N$-th root
of unity ($N$ the smallest integer such that $\varepsilon^N=1$)
and by $q$ in the general
case.

 The theories of chiral Potts \cite{CPI,CPII} type models, which saw
dramatic developments in recent years \cite{BA1,BA2,BA3,JI2,TA1}, are closely
tied to the representation theory of the quantum group $U_{\e}(sl(n,{\bf C}))$
in the case of $\e$ being an $N$-th root of unity. The progress in the
theories of chiral Potts models was partly stimulated by the better
understanding of its deep connection to
the representation theory of quantum groups.

 The  representation theory in the case of $\e$ an $N$-th
root of unity is much richer than for generic $q$, and several
deep results by  De Concini, Kac, Procesi \cite{KA1,KA2,KA3}
and Lusztig \cite{LU1,LU2,LU3,LU4} exist, laying the foundations
of the general representation theory in the roots of unity case.
Also considerable
progress has been made in directly constructing representations of
quantum groups $U_{\varepsilon}(g_n)$.
Accelerated by the development of chiral Potts type
models, much interest was devoted to find non-highest weight
representations of $U_{\varepsilon}(g_n)$
 \cite{RO,AR1,AR2,AR3,CH1,CH2,CH3}.  Finite
dimensional non-highest weight representations,
which do not exist in the representations theory
of $U_q(g_n)$,  are a new interesting feature of the representation theory
in the roots of unity case. Non-highest representations
of minimal dimension play a role similar to that played by
fundamental
representations of Lie algebras \cite{CH2}. The free parameters which are
characteristic of non-highest weight representations appear
in the form of spectral parameters in chiral Potts type models.

In this article we will show, that starting from an arbitrary
representation of $U_{\varepsilon}(g_{n-1})$ one can construct
a representation of $U_{\varepsilon}(g_n)$. The Lie algebras
$g_n$ and $g_{n-1}$ will usually, but not always lie in
the same series of Lie algebras.

De Concini, Kac \cite{KA1} showed that the maximal dimension
and number of parameters for representations of $U_{\varepsilon}(g_n)$
for odd $N$ is given by $N^{\Delta^{+}(g_n)}$ and dim$g_n$,
respectively.
In this expression
$\Delta^{+}(g_n)$ $(=1/2(\mbox{dim}g_n-n))$
is the number of positive roots of the
Lie algebra $g_n$.  We use here and in the following the term
dimension of a representation to denote the dimension of
its representation space.
The representations which will
be constructed in this article, are of
dimension greater than or equal to
$$N^{\Delta^{+}(g_n)-\Delta^{+}(g_{n-1})}.$$
 For $\varepsilon$-deformations in the case of the $A_n, B_n, C_n$ and
$D_n$ series such representations will be given in Section 3.
 The exceptional $E_6,E_7,E_8,F_4$ and $G_2$ cases of quantum groups are
discussed in Section 4. The number of free parameters in all constructed
representations is greater than or equal to $2(\Delta^{+}(g_n)-
\Delta^{+}(g_{n-1}))$. Representations of maximal dimension
and number of free parameters arise as special cases
for odd $N$ and are discussed in section 5. These representations
coincide with the maximal cyclic representations of Date et al.
\cite{JI3} in the $A_n$ case, and with the representations of \cite{SC}
in the $B_n, C_n$ and
$D_n$ cases. Conclusions are given in section 6 and in the appendix
two relations which are important for the construction
of representations in the case of $U_q(E_8)$ and $U_q(F_4)$ are written
down explicitly.

\sect{{ Definition of $U_q(g_n)$}}
In this article we will use the definition of
quantum groups given by the relations among
its Chevalley generators. The quantized
universal enveloping algebra $U_q(g_n)$ of a semi simple
Lie algebra $g_n$ of rank $n$ is generated by $4n$ Chevalley
generators $\{e_i,f_i,t_i^{\pm 1}\}$ which satisfy the commutation
relations
\begin{equation}\left.\begin{array}{rcl}
t_i t_j&=&t_j t_i, \quad t_i t_i^{-1}= t_i^{-1} t_i=1 \\
t_i e_j t_i^{-1}&=&q^{d_i a_{ij}} e_j, \quad
t_i f_j t_i^{-1}=q^{-d_i a_{ij}} f_j \label{def} \\
\left[e_i,f_i \right]&=&\delta_{ij}\left\{t_i\right\}_{q^{d_i}}
\end{array}\right\}
\end{equation}
and the Serre relations
\begin{equation}
\left.\begin{array}{rcl}
\sum\limits_{v=0}^{1-a_{ij}} (-1)^v\left[ \begin{array}{c}
1-a_{ij}\\
v
\end{array}\right]_{q^{d_i}}e_i^{1-a_{ij}-v} e_j e_i^v &=& 0 \\
\label{ser}\\
\sum\limits_{v=0}^{1-a_{ij}} (-1)^v\left[ \begin{array}{c}
1-a_{ij}\\
v
\end{array}\right]_{q^{d_i}}f_i^{1-a_{ij}-v} f_j f_i^v &=&0
\end{array}\right\}  (i\not= j).
\end{equation}
The matrix $a_{ij}$ is the Cartan matrix of the Lie algebra
$g_n$, $d_i$ non-zero integers satisfying $d_i a_{ij}=d_j a_{ji}$.
The $q$-Gaussian is given as
$$
\left[ \begin{array}{c}
m\\
n
\end{array}\right]_q = \frac{ [m]_q!}{[m-n]_q! [n]_q!}, \quad [m]_q!=
\prod_{j=1}^m
\frac{q^j-q^{-j}}{q-q^{-1}}\,
$$
and the curly bracket is defined by
$\{x\}_q=(x-x^{-1})/(q-q^{-1})$. Further, we extend the
algebra by adding the elements $t_i^{1/k}$, $k\in{\bf Z}$.
In case we specialize $q$ to be a primitive $N$-th root of unity the
letter $\varepsilon$ is used instead ($\varepsilon^N=1$). We will
not make use of the Hopf algebra structure of $U_q(g_n)$.

Some of the results in this article are most conveniently given
in terms of Weyl algebra generators. The
Weyl algebra $W$ is defined by the commutation relation among
its two generators $x,z$
\begin{equation}xz=qzx\quad . \end{equation}
We denote by $\tilde W$
a copy of this algebra with generators $\tilde x,\tilde z$.

 In the roots of unity case one can define a $N$ dimensional
representation
$\sigma_{gh}:W\to\mbox{End}({\bf C}^N)$ of $W$, depending on
the two parameters $g, h$, by
$$
\sigma_{gh}(x)=g\left(\begin{array}{ccccc}
0&1&0&\cdots&0\\
0&0&1&\cdots&0\\
\vdots&&&\ddots&\vdots\\
0&\cdots&&0&1\\
1&&\cdots&&0\\
\end{array}\right)
\quad
\sigma_{gh}(z)=h\left(\begin{array}{ccccc}
1&0&0&\cdots&0\\
0&\e&0&\cdots&0\\
0&0&\e^2&\cdots&0\\
&&&\ddots&\\
0&&\cdots&&\e^{N-1}\\
\end{array}\right).
$$
We will retain the notation $\sigma_{gh}$ also for representations
of tensor products of Weyl algebras $W$. In this cases
the letters  $g,h$ denote the set of parameters
$g:=\{g_1,\dots,g_l\}$ and $h:=\{h_1,\dots,h_l\}$, with $l$ the number
of $W$ algebras in  $\otimes_{1\leq i\leq l}W_i$.

\sect{{ The $A_n, B_n, C_n$ and $D_n$ series}}
In this section we want to demonstrate how one can construct, starting
from an arbitrary representation of a quantum group
$U_q(g_{n-1})$ a representation of $U_q(g_n)$, in the
cases of the $A_n, B_n, C_n$ and $D_n$ series of quantum groups.
The next section will be devoted to the more complicated
case of exceptional quantum groups.
As a first step we investigate the algebra $\bar U_q(g_{n-1})$
which is defined as the algebra given  by $U_q(g_n)$
generators $\F_i$,
($1\leq i\leq n-1$) and $\E_i,\T^{\pm 1}_i$, ($1\leq i\leq n$)
which satisfy the defining relations (\ref{def},\ref{ser}).
 Representations of algebras  $\bar U_q(g_{n-1})$ arise immedently from
representations of $U_q(g_{n-1})$, and in turn representations
of $U_q(g_n)$ itself will be defined in terms of
$\bar U_q(g_{n-1})$ representations. This will be discussed in the
following.

 The Cartan matrices in the $B_n$
and $C_n$ correspond to Dynkin diagrams with the first root
being the shortest or longest root, respectively. For the $D_n$ case
we fix the notation in a way  such that the Dynkin diagram
nodes $1$ and $2$ are both connected with node $3$. The integers
$d_i=1$ for $2\leq i\leq n$ in all cases and $d_1=1,1/2,2,1$ in the
$A_n$, $B_n$, $C_n$ and $D_n$ series, respectively.

\begin{lemma}\label{uhat}
Let $\{ f_i,  e_i,  t_i,  t_i^{-1}\}$ $(i=1,\dots, n-1)$
be the generators of $U_q(g_{n-1})$ and
$\{\E_i, \T_i,  \T_i^{-1}\}$ $(i=1,\dots, n)$,
$\{ \F_i\}$ $(i=1,\dots,n-1)$ the generators of $\bar U_q(g_{n-1})$.
Then one obtains an algebra homomorphism $\bar\rho$ from $\bar U_q(g_{n-1})$
to $U_q(g_{n-1})$
by taking $\bar\rho(\F_i) =f_i$, $\bar\rho(\E_i) =e_i$, $\bar\rho(\T^{\pm 1}_i)
=t^{\pm 1}_i$ $(i=1,\dots,n-1)$ and $\bar\rho(\E_{n})=0$. The
action of the algebra homomorphism $\bar\rho$ on the generators
$ \T_{n}^{\pm 1}$
depends on the complex parameters $\lambda_n$ and is defined by
$$
\bar\rho( \T_{n})=
\left\{
\begin{array}{lc} q^{\lambda_n}
\prod\limits_{i=1}^{n-1} t_i^{-\frac{i}{n}}& \quad\mbox{for}\quad\bar
U_q(A_{n-1})\\
q^{\lambda_n}\prod\limits_{i=1}^{n-1} t_i^{-1}& \quad\mbox{for}\quad\bar
U_q(B_{n-1})\\
q^{\lambda_n}t_1^{-\frac{1}{2}}
\prod\limits_{i=2}^{n-1} t_i^{-1}& \quad\mbox{for}\quad
\bar U_q(C_{n-1})\\
q^{\lambda_n}t_1^{-\frac{1}{2}}t_2^{-\frac{1}{2}}
\prod\limits_{i=3}^{n-1} t_i^{-1}& \quad\mbox{for}\quad
\bar U_q(D_{n-1})
\end{array}
\right.
$$
\end{lemma}

The above algebra homomorphism $\bar\rho$ can be
extended to define
representations of the quantum group $U_q(g_n)$.
The above lemma, together with the theorem below
shows how any representation of
$U_q(g_{n-1})$ gives rise to a representation of $U_q(g_{n})$
in the $A_n, B_n, C_n$ and $D_n$ series.

\begin{theorem}\label{t1}
The following formulas define algebra homomorphisms $\rho$
from the quantum
groups $U_q(g_n)$ to $(\otimes_{i} W)\otimes \bar U_q(g_{n-1})$,
for $g_n$ the $A_n, B_n, C_n$ and $D_n$
series of semi simple Lie algebras. The composition
$\pi:=(\sigma_{gh}\otimes \mbox{id})\cdot \rho$ defines
an algebra homomorphism in the roots of unity case.
$\gdef\so{{\scriptscriptstyle\otimes}}$
$\gdef\sv{\underline}$$\gdef\bx{\tilde x}$$\gdef\bz{\tilde z}$
\\
\begin{eqnarray*}
\bf a.)&& \rho:U_q(A_n)\to (\ot\limits_{i=1}^n W_i)\otimes
\bar U_q(A_{n-1})\\
\rho(f_i)&=&\{z_{i-1}z_i^{-1}\}x_i +x_{i-1}^{-1}x_i  \F_{i-1},
\quad \rho(e_i)=\{z_iz_{i+1}^{-1}{ \T_i}\}x_i^{-1}
+{ \E_i}\\
\rho(t_i)&=&z_{i-1}^{-1}z_i^2z_{i+1}^{-1}{ \T_i}\\&&\\
\bf b.)&& \rho:U_q(B_n)\to (\ot\limits_{i=1}^n W_i)\ot
(\ot\limits_{i=1}^{n-1}{\tilde W_i})\otimes \bar U_q(B_{n-1})\\
\rho(f_i)&=&\{z_{i+1}z_{i}^{-1}\}x_i+
\{\bz_{{ i}}\bz_{{ i+1}}^{-1}\}x_{i+1}^{-1}x_i\bx_{{i+1}}
+x_{i+1}^{-1}x_i\bx_{{ i+1}}\bx_{{i}}^{-1} \F_i,\quad (i>1)\\
\rho(f_1)&=&\{z_2z_1^{-{1/2}}\}_{q^{{1/2}}}x_1+
\{\bz_{{2}}^{-1}z_1^{{1/2}}\}_{q^{{1/2}}}x_2^{-1}x_1\bx_{{2}}
+x_2^{-1}\bx_{{2}} \F_1\\
\rho(e_i)&=&\{z_iz_{i-1}^{-1}\bz_{{ i-1}}^{-1}\bz_i^2\bz_{i+1}^{-1}
\T_i\}x_i^{-1}+\{\bz_i\bz_{{ i+1}}^{-1}
 \T_i\}
\bx^{-1}_{{ i}}+ \E_i,\quad (i>1)\\
\rho(e_1)&=&\{z_1^{{1/2}}\bz_2^{-{1}} \T_1\}_{q^{{1/2}}}
x_1^{-1}+ \E_1\\
\rho(t_i)&=&z_{i+1}^{-1}z_i^2z_{i-1}^{-1}\bz_{{ i-1}}^{-1}\bz_{{ i}}^2
\bz_{{ i+1}}^{-1} \T_i, \quad (i>1),\quad
\rho(t_1)=z_2^{-{1}}z_1\bz_{{2}}^{-{1}} \T_1
\\&&\\
\bf c.)&& \rho:U_q(C_n)\to (\ot\limits_{i=1}^n W_i)\ot
(\ot\limits_{i=1}^{n-1}{\tilde W_i})\otimes \bar U_q(C_{n-1})
\\
&&\mbox{$\rho(f_i), \rho(e_i), \rho(t_i)$ as in the $B_n$ case $(i>1)$}\\
\rho(f_1)&=&\{z_1^2\bz_{{2}}^{-2}\}_{q^2}x_2^{-2}x_1\bx_{{2}}^2
+ \{z_{2}\bz^{-1}_{{2}}\}x_2^{-1}x_1\bx_{{2}}+
\{z_{2}^2z_1^{-2}\}_{q^2}x_1+x_{2}^{-2}\bx_{{2}}^2 \F_1\\
\rho(e_1)&=&\{z_1^2\bz_2^{-2} \T_1\}_{q^2}x_1^{-1}
+ \E_1, \quad
\rho(t_1)=z_2^{-2}z_1^4\bz_{{2}}^{-2} \T_1
\\&&\\
\bf d.)&& \rho:U_q(D_n)\to (\ot\limits_{i=1}^n W_i)\ot
(\ot\limits_{i=1}^{n-2}{\tilde W_i})\otimes \bar U_q(D_{n-1})
\\
&&\mbox{$\rho(f_i), \rho(e_i), \rho(t_i)$ as in the $B_n$ case $(i>2)$}\\
\rho(f_1)&=&\{z_3z_1^{-1}\}x_1+
\{z_{2}\bz_{{3}}^{-1}\}\bx_{{3}}
x_3^{-1}x_1+x_{3}^{-1}x_1x_2^{-1}\bx_{{3}} \F_2\\
\rho(f_2)&=&\{z_3z_2^{-1}\}x_{2}+\{z_1\bz_{{3}}^{-1}\}
\bx_{{3}}x_{3}^{-1}x_2+
x_3^{-1}x_2x_1^{-1}\bx_{{3}} \F_1\\
\rho(e_1)&=&\{z_1\bz_{{3}}^{-1}
 \T_1\}x_1^{-1}+ \E_1,\quad
\rho(e_2)=\{z_2\bz_{{3}}^{-1}
 \T_2\}x_{2}^{-1}+ \E_2\\
\rho(t_1)&=&z_3^{-1}z_1^2\bz_{{3}}^{-1} \T_1, \quad
\rho(t_2)=z_3^{-1}z_2^2\bz_{{3}}^{-1} \T_2\quad.
\end{eqnarray*}
In these formulas expressions of type
$x_i\otimes \F_j$ were abbreviated writing $x_i \F_j$.
Further, we set $\F_0=0$ and
$x_i^{-1}= \bx_i^{-1}=0, z_i=\bz_i=1$ if the index $i$ is out of range.

Taking $\bar U_{\e}(g_{n-1})$ in an arbitrary  representation $\bar\rho'$,
then the algebra homomorphisms $\pi$ becomes a representation $\pi'$
of dimension
$N^{\Delta^{+}(g_n)-\Delta^{+}(g_{n-1})}$ times the dimension
of $\bar\rho'$.
The number of free parameters in $\pi'$ becomes
$2(\Delta^{+}(g_n)-\Delta^{+}(g_{n-1}))$ plus the number of free
parameters in $\bar\rho'$.
\end{theorem}

This theorem can be proven by directly verifying the defining relations
of the algebras. We omit the details of these calculations.
\begin{remark}\rm
Representations with dimensions equal to
$N^{\Delta^{+}(g_n)-\Delta^{+}(g_{n-1})}$ and
$2(\Delta^{+}(g_n)-\Delta^{+}(g_{n-1})+\frac{1}{2})$ free parameters are
obtained by taking the trivial representation $\rho'_0$ of $U_{\e}(g_{n-1})$
($\rho'_0(e_i)=0, \rho'_0(f_i)=0, \rho'_0(t_i)=1,
\, 1\leq i\leq n-1$) to define $\bar \rho$
in lemma \ref{uhat}. In the case of $U_{\e}(A_n)$ the resulting
 representations $\pi'$
are minimal cyclic representations \cite{AR1,AR2,CH1,CH2,JI2}.
Further representations of
dimension $N^{\Delta^{+}(g_n)-\Delta^{+}(g_{n-1})}$ were discussed
in the quantum $SO(5)$ case in \cite{AR2} and for quantum $SO(8)$
in \cite{CH3}.
\end{remark}

\begin{remark}\rm
The algebra homomorphism $\pi$ defined above for the $A_n$ series is
cyclic in the sense that the Chevalley operators in the algebra homomorphism
$\pi$
to the power of $N$ are non-vanishing scalars for generic values
of parameters. The explicit expressions are given
in the following formulas.
\begin{eqnarray*}
\pi(f_i)^N&=&-\frac{h_{i-1}^Nh_i^{-N}(-1)^N+h_{i-1}^{-N}h_i^{N}}{(q-q^{-1})^N}
g_i^N+g_{i-1}^{-N}g_i^N \F_{i-1}^N\\
\pi(e_i)^N&=&-\frac{h_{i}^Nh_{i+1}^{-N}
 \T_i^{N}(-1)^N+h_{i}^{-N}h_{i+1}^{N} \T_i^{-N}}
{(q-q^{-1})^N}
g_{i+1}^{-N}+ \E_{i}^N\\
\pi(t_i)&=&h_{i-1}^{-N}h_i^{2N}h_{i+1}^{-N} \T_i^N\quad.
\end{eqnarray*}
These formulas also show that
restricted representations ($\pi'(f_i)^N=0, \pi'(e_i)^N=0,
\pi'(k_i)^N=1$), as well as semi-cyclic representations
(either $\pi'(f_i)^N=0$ or $ \pi'(e_i)^N=0$) can be derived from
the cyclic representation by specializing
some or all of the free parameters. Cyclic and semi-cyclic
representations of $U_{\e}(A_n)$ in dimensions higher than the minimal
dimensions were discussed in \cite{AR1,JI4}.
\end{remark}

Theorem \ref{t1} allows to construct a representation of
$U_q(g_n)$ from representations of
smaller rank quantum groups $U_q(g_{n-1})$ with
$g_n$ and $g_{n-1}$ being Lie algebras of the same series.
To show that they need not to be necessarily from the same series
we present as an example how a representation of
$U_q(C_3)$ follows from a representation of
$U_q(A_2)$. We define  $\bar U_q'(A_2)$ to be
the algebra given by the generators $\E_i,\T_i^{\pm1}$, ($i=1,2,3$)
and $\F_2,\F_3$ of a $U_q(C_3)$ algebra.
Similarly to \ref{uhat} one can give an
algebra homomorphism $\bar\rho$ from $\bar U_q'(A_2)$ to
$U_q(A_2)$ by setting
$\bar\rho(\E_1)=0$,
$ \bar\rho(\T_1)=q^{\lambda_1}
t_3^{-2/3}t_2^{-4/3}$ and the other
generators equal to $U_q(A_2)$ generators.

\begin{example}\rm
The map $\rho: U_q(C_3)\to (\ot_{i=1}^6 W_i)\otimes \bar U_q'(A_2)$
given below, defines an algebra homomorphism. The
composition
$\pi:=(\sigma_{gh}\otimes \mbox{id})\cdot\rho$ gives
 an algebra homomorphism in the roots of unity case.
Taking $\bar U_{\e}'(A_2)$ in an arbitrary representation $\bar\rho'$ one
obtains a representation $\pi'$ of $U_{\e}(C_3)$
with dimension $N^6$ times the dimension of $\bar\rho'$.
 Twelve free parameters arise from taking
representation $\sigma_{gh}$ of $(\ot_{i=1}^6 W_i)$ and further
free parameters can arise in the representation $\bar\rho'$.
\begin{eqnarray*}
\rho(f_1)&=&\{z_1^{-2}\}_{q^2}x_1\\
\rho(f_2)&=&\{z_2z_4^{-2}\}x_1^{-1}x_2x_4+
\{z_3z_5^{-1}\}x_1^{-1}x_4x_5+
\{z_1^2z_2^{-1}\}x_2+
x_1^{-1}x_3^{-1}x_4x_5  \F_3\\
\rho(f_3)&=&\{z_5z_6^{-2}\}x_2^{-1}x_3x_4^{-1}x_5x_6+
\{z_4^2z_5^{-1}\}x_2^{-1}x_3x_5 +
\{z_2z_3^{-1}\}x_3+
x_2^{-1}x_3x_4^{-1}x_6  \F_2\\
\rho(e_1)&=&\{z_1^2z_2^{-2}z_4^4z_5^{-2}z_6^4 \T_1\}_{q^2}x_1^{-1}+
\{z_4^2z_5^{-2}z_6^4 \T_1\}_{q^2}x_4^{-1}+
\{z_6^2 \T_1\}_{q^2}x_6^{-1}+\E_1\\
\rho(e_2)&=&\{z_2z_3^{-1}z_4^{-2}z_5^2z_6^{-2}\T_2\}x_2^{-1}+
\{z_5z_6^{-2} \T_2\}x_5^{-1}+
\E_2\\
\rho(e_3)&=&\{z_3z_5^{-1} \T_3\}x_3^{-1}+
 \E_3\\
\rho(t_1)&=&z_1^{4}z_2^{-2}z_4^4z_5^{-2}z_6^4 \T_1\\
\rho(t_2)&=&z_1^{-2}z_2^2z_3^{-1}z_4^{-2}z_5^2z_6^{-2} \T_2\\
\rho(t_3)&=&z_2^{-1}z_3^2z_5^{-1} \T_3
\end{eqnarray*}
\end{example}

\sect{{The exceptional $E_6, E_7, E_8, F_4$ and $G_2$ quantum
groups}}
So far we did not describe the actual construction method which
leads to the algebra homomorphisms which were given
in the previous section.
The  procedure will be outlined in the following
in the cases of the exceptional quantum Lie algebras and
the $A_n$ case will appear as an example.

The construction method is based on a set of relations
in $U_q(g_n)$. The relations needed the derive algebra
homomorphisms of quantum Lie algebras in the $E_6,E_7,E_8$ and
$F_4$ cases are given in the following. We denote
divided powers of $f_i$ generators ${f^{j}_i}/{\lbrack j \rbrack!}$
by $f_i^{(j)}.$
$\gdef\[{\lbrack}$
$\gdef\]{\rbrack}$

\begin{definition}\label{d1}\rm
In $U_q(g_n)$ we have the commutation relations (compare \cite{JI3,SC})
\begin{eqnarray*}
(i) &&f_i f^{(j)}_i=\[j +1\]_{q^{d_i}}f_i^{(j+1)}\\
(ii) &&f_if^{(j_1)}_kf_i^{(j_2)}=
f_k^{(j_1)}f_i^{(j_2+1)}\[-j_1+j_2+1\]_{q^{d_i}}
+f^{(j_1-1)}_kf_i^{(j_2+1)}f_k,\\&& \mbox{if $a_{ik}=a_{ki}=-1$}\\
(iii)&&f_if_k^{(j_1)}f_i^{(j_2)}f_k^{(j_3)}=
f_k^{(j_1-2)}f_i^{(j_2+1)}f_k^{(j_3+2)}\[-j_2+j_3+1\]_{q^2}+\\&&
f_k^{(j_1-1)}f_i^{(j_2+1)}f_k^{(j_3+1)}\[-j_1+j_2+2\]+
f_k^{(j_1)}f_i^{(j_2+1)}f_k^{(j_3)}\[-j_1+j_2+1\]_{q^2}+\\&&
f_k^{(j_1-2)}f_i^{(j_2)}f_k^{(j_3+2)}f_i,\mbox{ if $a_{ki}=-2, a_{ik}=-1$}\\
(iv)&&e_if_i^{(j)}=f_i^{(j)}e_i+f_i^{(j-1)}\{q^{d_i(1-j)}t_i\}_{q^{d_i}}
\\
(v)&& t_if_k^{(j)}=q^{-j a_{ik}d_i }f_k^{(j)}t_i\\
(vi)&&  f_i f_k^{(j)}=f_k^{(j)}f_i \mbox{ and }
f_i^{(j_1)}f_k^{(j_2)}=f_k^{(j_2)}f_i^{(j_1)} \mbox{ if }
a_{ik}=0, \\&& e_i f_k^{(j)}=f_k^{(j)}e_i, \mbox{ if } i\not=k\\
(vii)&& \mbox{relation in lemma \ref{l1}, appendix
(used only in the $E_8$ case)}\\
(viii)&& \mbox{relation in lemma \ref{l2}, appendix
(used only in the $F_4$ case)}
\end{eqnarray*}
\end{definition}

We will use the above relations to commute single generators
$f_i,e_i$ and $t_i$ through monomials of $l$ factors  of type
\begin{equation}f_{i_1}^{(j_1)}\cdots f_{i_l}^{(j_l)}
\label{mono}\end{equation}
($l=\Delta^{+}(g_n)-
\Delta^{+}(g_{n-1})$).
We say a generator $f_i,e_i$ or $t_i$ commutes with such a
monomial if the multiplication of the generator on the monomial
from the left gives a sum over monomials of the same type multiplied
by a single or none generator from the right. In this sense
we say that $f_i$ commutes with the monomial $f_k^{(j_1)}f_i^{(j_2)}$
according to relation $(iii)$ in \ref{d1}, if $a_{ik}=a_{ki}=-1$.
One can find monomials with $l$ factors
for all quantum Lie algebras $U_q(g_n)$ which commute with all of
its generators. Moreover, the relations \ref{d1} will be sufficient
to commute the generators of $U_q(E_6), U_q(E_7),U_q(E_8)$ and $U_q(F_4)$
with monomials in the corresponding algebras.
The $U_q(G_2)$ case will be treated separately.

Commuting a generator $s$ $\in\{f_i,e_i,t_i\}$
with a monomial of type (\ref{mono})
using exclusively the relations \ref{d1} we denote by rel($sf_{i_1}^{(j_1)}
\cdots f_{i_l}^{(j_l)}$).
To apply relations \ref{d1} in this way will be
the first step in a construction procedure which leads
to representations of $U_q(g_n)$ in the exceptional
cases.

The second step introduces the Weyl algebra generators.
We shall give a rule $\Omega$  which applies on expressions of the following
type
\begin{equation}
q^{r_1j_1+\cdots+r_l j_l}f_{i_1}^{(j_1+\alpha_1)}\cdots
f_{i_l}^{(j_l+\alpha_l)}s\quad,
\label{exp}\end{equation}
wherein the integer quantities $j_1,\cdots,j_l$ are regarded as
``free variables'' and $r_k,\alpha_k\in {\bf N},
1\leq k\leq l$.
The factor $s$ is
either $1$ or a single $U_q(g_n)$ generator $f_m,e_m, t^{\pm 1}_m$,
($1\leq m\leq n$). If $a,b$ are two expressions of type (\ref{exp})
and $\beta$ a rational function in $q$
which does not depend on the $j_k$'s,
then $\Omega$ satisfies
$\Omega(a+ b)=\Omega(a)+\Omega(b)$ and
 $\Omega(\beta a)=\beta \Omega(a)$.
The rule $\Omega$ applies on expressions (\ref{exp})
according to the following assignment
\begin{equation}
\Omega: q^{r_1j_1+\cdots+r_l j_l}f_{i_1}^{(j_1+\alpha_1)}
\cdots f_{i_l}^{(j_l+\alpha_l)}s\mapsto x_1^{\alpha_1}z_1^{-r_1}\cdots
x_l^{\alpha_l}z_l^{-r_l}S \quad.
\label{om}\end{equation}
If $s=1,f_m,e_m,t^{\pm1}_m$ then $S=1,\F_m,
\E_m,\T^{\pm1}_m$, respectively.
Capital $\E_i, \F_i$ and
$\T_i^{\pm1}$ are  generators of a second $U_q(g_n)$ algebra, and will
generate in the coming examples the algebras $\bar U_q(g_{n-1})$.

To illustrate the working of the above two steps in the construction
procedure of an algebra homomorphism we consider the $A_n$ case of section 3.

\begin{example}\label{e2}\rm
The expressions for the
algebra homomorphism $\rho$ from $U_q(A_n)$ to
$W_1\otimes \cdots \otimes W_n\otimes
\bar U_q(A_{n-1})$ in theorem \ref{t1} can
be constructed in a unique way using relations \ref{d1} and $\Omega$.
If one defines $y$ to be
 $$y=f_1^{(j_1)}f_2^{(j_2)}\cdots f_n^{(j_n)}$$ then the
formula for $\rho(f_i)$ in theorem
\ref{t1}, a.) is given by $\Omega(\mbox{\rm rel}(f_i y))$,
using relations
$(vi, ii)$ in \ref{d1} and  (\ref{om}).
Similarly, one obtains $\rho(e_i)=\Omega(\mbox{\rm rel}(e_i y))$
and $\rho(t_i)=\Omega(\mbox{\rm rel}(t_i y))$,
using the relations $(iv, v, vi)$ in \ref{d1} and (\ref{om}).
The monomial $y$ which is used to construct the algebra homomorphism
$\rho$  in the $U_q(B_n)$ and $U_q(C_n)$ case (theorem \ref{t1}, b.), c.))
is
$$
y=f_n^{(j_1)}f_{n-1}^{(j_2)}\cdots f_2^{(j_{n-1})} f_1^{(j_n)}f_2^{(j_{n+1})}
\cdots f_n^{(j_{2n-1})}\, ,
$$
and a similar monomial
$$
y=f_n^{(j_1)}f_{n-1}^{(j_2)}\cdots f_2^{(j_{n-1})} f_1^{(j_n)}f_3^{(j_{n+1})}
\cdots f_n^{(j_{2n-2})}\, ,
$$
is used to derive the algebra homomorphism
in the $U_q(D_n)$ case.
\end{example}

We start the investigation of the exceptional quantum Lie
algebras with the $U_q(E_6), U_q(E_7), U_q(E_8)$ cases
and define the following numbering of the nodes
in the corresponding Dynkin diagrams
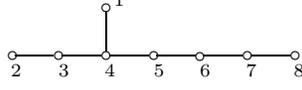
\begin{center}
{\tt    \setlength{\unitlength}{0.5pt}
\begin{picture}(242,83)
\thinlines    \put(15,26){\line(1,0){29}}
              \put(13,26){\circle{6}}
              \put(48,26){\circle{6}}
              \put(51,26){\line(1,0){29}}
              \put(84,26){\circle{6}}
              \put(87,26){\line(1,0){29}}
              \put(120,26){\circle{6}}
              \put(122,26){\line(1,0){29}}
              \put(158,26){\line(1,0){29}}
              \put(191,26){\circle{6}}
              \put(155,25){\circle{6}}
              \put(194,26){\line(1,0){29}}
              \put(227,26){\circle{6}}
              \put(84,29){\line(0,1){30}}
              \put(84,62){\circle{6}}
              \put(90,63){\smmm 1}
              \put(155,10){\smmm 6}
              \put(190,10){\smmm 7}
              \put(226,10){\smmm 8}
              \put(12,10){\smmm 2}
              \put(48,10){\smmm 3}
              \put(83,10){\smmm 4}
              \put(120,10){\smmm 5}
\end{picture}}
\end{center}
The integers $d_i$ are equal to $1$ for all $i$.
Let us introduce the algebra $\bar U_q'(D_5)$ being generated
by  $\F_{i\not=2}, \E_i, \T_i^{\pm1}\in$  $U_q(E_6)$. Similarly,
one can define $\bar U_q(E_6)$
as $\F_{i\not=7}, \E_i, \T_i^{\pm1}\in U_q(E_7)$ and
$\bar U_q(E_7)$ as given by the generators
$\F_{i\not=8}, \E_i, \T_i^{\pm1}\in U_q(E_8)$.
Algebra homomorphism $\bar\rho$ in case of these algebras arise analogously
 as in lemma \ref{uhat}
by taking
\begin{eqnarray*}
\bar\rho(\T_2)&=&q^{\lambda_2}t_1^{-\frac{3}{4}} t_3^{-\frac{5}{4}}
t_4^{-\frac{3}{2}}t_5^{-1}t_6^{-\frac{1}{2}},\\
\bar\rho(\T_7)&=&q^{\lambda_7}t_1^{-1}t_2^{-\frac{2}{3}}t_3^{-\frac{4}{3}}
t_4^{-2}t_5^{-\frac{5}{3}}t_6^{-\frac{4}{3}},\\
\bar\rho(\T_8)&=&q^{\lambda_8}t_1^{-\frac{3}{2}}t_2^{-1}t_3^{-2}
t_4^{-3}t_5^{-\frac{5}{2}}t_6^{-2}t_7^{-\frac{3}{2}},
\end{eqnarray*}
respectively.

Let us abbreviate monomials of generators in $U_q(g_n)$ of type
$$f_i^{(j_k)}f_{i\pm1}^{(j_{k+1})}\cdots f_{i\pm l\mp1}^{(j_{k+ l-1})}
f_{i\pm l}^{(j_{k+l})}$$
by $f^{(j_k)}_{i i\pm l}$. Using this notation we
define the following monomials in $U_q(E_6)$, $U_q(E_7)$ and $U_q(E_8)$
\begin{equation}
\begin{array}{lcl}
y_6&=&f_{63}^{(j_1)}f_1^{(j_5)}f_{46}^{(j_{6})}  f_{25}^{(j_{9})}
f_1^{(j_{13})}  f_{42}^{(j_{14})}\\
y_7&=& f_{73}^{(j_{1})}  f_1^{(j_{6})}  f_{47}^{(j_{7})}
f_{26}^{(j_{11})}  f_1^{(j_{16})}  f_{45}^{(j_{17})}  f_{34}^{(j_{19})}
  f_{23}^{(j_{21})}  f_1^{(j_{23})} f_{47}^{(j_{24})}\\
y_8&=&f_{83}^{(j_{1})}f_{1}^{(j_{7})}f_{48}^{(j_{8})}f_{27}^{(j_{13})}
f_{1}^{(j_{19})}f_{46}^{(j_{20})}f_{35}^{(j_{23})}f_{24}^{(j_{26})}
f_{1}^{(j_{29})}f_{42}^{(j_{30})} f_{53}^{(j_{33})}f_{64}^{(j_{36})}
f_{1}^{(j_{39})}\\&&
f_{72}^{(j_{40})}f_{84}^{(j_{46})}f_{1}^{(j_{51})}f_{38}^{(j_{52})}
\end{array}
\label{wais}
\end{equation}

Similar to example \ref{e2} one can obtain the expressions for
algebra homomorphisms in the case
of $U_{\e}(E_6),U_{\e}(E_7)$ and $U_{\e}(E_8)$
by using relations \ref{d1} and $\Omega$.

\begin{theorem}
The following expressions define an algebra homomorphism $\rho$
in the case of the quantum
groups $U_q(E_6)$, $U_q(E_7)$ and $U_q(E_8)$.
The composition $\pi:=(\sigma_{gh}\otimes \mbox{id})\cdot\rho$ defines
an algebra homomorphisms in the roots of unity case.
 The mappings $\rho$ are defined by
\begin{eqnarray*}
{\bf e6.)}&&  \rho:U_q(E_6)\to (\ot\limits_{k=1}^{16} W_k)\otimes
\bar U_q'(D_5),\\
{\bf e7.)}&&  \rho:U_q(E_7)\to (\ot\limits_{k=1}^{27} W_k)\otimes
\bar U_q(E_6),\\
{\bf e8.)}&&  \rho:U_q(E_8)\to (\ot\limits_{k=1}^{57} W_k)\otimes
\bar U_q(E_7),
\\&&\rho(f_i)=\Omega(\mbox{\rm rel}(f_i y_n)),\quad
\rho(e_i)=\Omega(\mbox{\rm rel}(e_i y_n)),\\&&
\rho(t_i)=\Omega(\mbox{\rm rel}(t_i y_n)),\quad \forall i,\mbox{ and }n=6,7,8
\end{eqnarray*}
Taking $\bar U_{\e}'(D_5), \bar U_{\e}(E_6), \bar U_{\e}(E_7)$ in an
arbitrary representation $\bar\rho'$, one obtains representations $\pi'$
which dimensions are given respectively by $N^{16}$, $N^{27}$
and $N^{57}$ times the dimensions of $\bar\rho'$.
The number of free parameters of the representations $\pi'$ is
$32$, $54$ and $114$, respectively plus the number of free parameters
in a representation $\bar\rho'$.
\end{theorem}

The remaining two cases of exceptional Lie algebras are discussed
in the following theorems. We fix the
numbering of nodes in the Dynkin diagram
for the $F_4$ case as
\begin{center}

{\tt    \setlength{\unitlength}{0.5pt}
\begin{picture}(163,50)
\thinlines
              \put(147,13){\smmm 4}
              \put(102,13){\smmm 3}
              \put(58,12){\smmm 2}
              \put(13,13){\smmm 1}
              \put(147,26){\circle{8}}
              \put(14,26){\circle{8}}
              \put(17,27){\line(1,0){36}}
              \put(106,27){\line(1,0){36}}
              \put(102,27){\circle{8}}
              \put(58,27){\circle{8}}
              \put(61,26){\line(1,0){36}}
              \put(62,29){\line(1,0){36}}
              \put(89,20.5){>}
\end{picture}}
\end{center}
The integers $d_i$ are defined as $d_1=d_2=2, d_3=d_4=1$.
We write $\bar U_q'(B_3)$ for the
algebra generated by $\F_{i\not= 4},\E_i,\T_i^{\pm1}$
$\in U_q(F_4)$.
Similarly to lemma \ref{uhat} one can give
an algebra homomorphism $\bar\rho$  in the case of
 $\bar U_q'(B_3)$ starting from
$U_q(B_3)$ and
defining $\bar\rho(\T_4)=q^{\lambda_4}t_1^{-1/2}t_2^{-1}t_3^{-3/2}$.
Using of the relations \ref{d1} and the operation $\Omega$
in the $U_{\e}(F_4)$ case, one can again
define the algebra homomorphism in a short way. Let $y_4$ denote the monomial
$y_4=f_{14}^{(j_1)}f_{2}^{(j_{5})}f_{31}^{(j_{6})}
f_{24}^{(j_{9})}f_{2}^{(j_{12})}f_{31}^{(j_{13})}$.

\begin{theorem}\label{tf4}
The following expressions define an algebra homomorphism
$\rho$ from  $U_q(F_4)$ to
$W_1\ot\cdots W_{15}\ot \bar U_q'(B_3)$
in a unique way and by composition
with $\s_{gh}$ an algebra homomorphism in the roots of unity case.
\begin{eqnarray*}{\bf f.)}&&
\rho: U_q(F_4)\to (\ot\limits_{i=1}^{15} W_i)\ot \bar U_q'(B_3)\\
&&\rho(f_i)=\Omega(\mbox{\rm rel}(f_i y_4)), \quad
\rho(e_i)=\Omega(\mbox{\rm rel}(e_i y_4)), \quad \rho(t_i)=
\Omega(\mbox{\rm rel}(t_i y_4)),\quad \forall i
\end{eqnarray*}
Taking $\bar U_q'(B_3)$ in an arbitrary representation $\bar\rho'$
one obtains a representation $\pi'$ of dimension $N^{15}$
times the dimension in $\bar\rho'$.
The number of free parameters in $\pi'$ is $30$ plus the number of free
parameters in $\bar\rho'$.
\end{theorem}

Let $G_2$ be defined by the Cartan matrix with $a_{11}=2,a_{12}=-3, a_{21}=-1$
and $a_{22}=2$. The integers $d_i$ are given as $d_1=1,d_2=3$.
In the following $\bar U_q'(A_1)$ is defined by the $U_q(G_2)$ generators
$\F_1,\E_1,\E_2,\T_1^{\pm1},\T_2^{\pm1}$.
In analogy to lemma \ref{uhat} one can easily give
a corresponding  algebra homomorphism $\bar \rho$,  starting
from $U_q'(A_1)$ and taking $\bar\rho (\T_2)=q^{\lambda_2}t_1^{-3/2}$.
\begin{theorem}\label{t8}
The following formulas define an algebra homomorphism $\rho$ from $U_q(G_2)$
to $W_1\ot\cdots\ot W_5\ot \bar U_q'(A_1)$
and by composition with $\s_{gh}$ an algebra homomorphism
$\pi:=(\sigma_{gh}\otimes \mbox{id})\cdot\rho$ in the
roots of unity case.
\begin{eqnarray*}{\bf g.)}&&
\rho: U_q(G_2)\to (\ot\limits_{i=1}^{5} W_i)\ot \bar U_q'(A_1)\\
\rho(f_1)&=&\{z_3^3z_4^{-2}\}x_1^{-1}x_2^{-1}x_3x_4^2+
\{z_4z_5^{-3}\}x_1^{-1}x_2^{-1}x_4^2x_5+
\{x_2^2x_3^{-3}\}x_1^{-1}x_2x_3+\\&&
\{z_2z_4^{-1}\}\{q^2\}x_1^{-1}x_3x_4+
\{z_1^3z_2^{-1}\}x_2+x_1^{-1}x_2^{-1}x_4x_5 \F_1\\
\rho(f_2)&=&\{z_1^{-3}\}_{q^3}x_1\\
\rho(e_1)&=&\{z_2z_3^{-3}z_5^{-3}z_4^2 \T_1\}x_2^{-1}+
\{z_4z_5^{-3} \T_1\}x_4^{-1}+ \E_1\\
\rho(e_2)&=&\{z_1^{3}z_3^6z_5^6z_2^{-3}z_4^{-3} \T_2\}_{q^3}x_1^{-1}+
\{z_3^3z_5^6z_4^{-3}\T_2\}_{q^3}x_3^{-1}+
\{z_5^3 \T_2\}_{q^3}x_5^{-1} +  \E_2\\
\rho(t_1)&=&z_1^{-3}z_3^{-3}z_5^{-3}z_2^2z_4^2 \T_1\\
\rho(t_2)&=&z_1^6z_3^6z_5^6z_2^{-3}z_4^{-3} \T_2
\end{eqnarray*}
Taking $\bar U_{\e}'(A_1)$ in an arbitrary representation $\bar\rho'$
one obtains a representation $\pi'$ which
dimension (number of free parameters) is $N^{5}$
(10) times (plus) the dimension (number of free parameters)
of $\bar\rho'$, respectively.
\end{theorem}

\sect{{ Representations of maximal dimensions}}
The representations constructed in section 3 and 4 depend on the arbitrary
representations $\bar\rho'$
of the corresponding $\bar U_{\e}(g_{n-1})$ algebras.
In this section we want to show how a natural choice of this
representation of $\bar U_{\e}(g_{n-1})$ leads to representations
of $U_{\e}(g_{n})$, which have  maximal dimensions and number of parameters
for odd $N$.

Irreducible representations of quantum groups $U_{\e}(g_n)$ in the
roots of unity case exist only in dimensions smaller than or equal
to $N^{\Delta^{+}(g_n)}$ and the number
of free parameters is maximally dim$(g_n)$, for odd $N$ \cite{KA1}.
The simplest representation of this type is the non-highest
weight representation $\pi':=\sigma_{gh}\cdot\rho$ of $U_{\e}(A_1)$ which is
defined in terms of Weyl algebra generators by
$$
\rho(f)=\{z^{-1}\} x,\quad\rho(e)=\{\e^{\lambda_1}z\} x^{-1},\quad
\rho(t)=\e^{\lambda_1}z^{2}\, .
$$
Representation $\pi'$ has dimension $N$ and the map $\s_{gh}$
together with the complex parameter $\lambda_1$ gives rise
to the  $3$ free parameters of the representation.
Starting from this representation one can step by step
construct representations of higher rank Lie algebras,
using inductively representations $\bar\rho$ of $\bar U_{\e}(g_{n-1})$ and
representations $\pi'$ of $U_{\e}(g_n)$.
A representation of $U_{\e}(g_n)$ obtained in this way will
have  $2\Delta^{+}(g_n)$ number of free parameters induced by the
free parameters of the mappings $\s_{gh}$
and in  addition $n$ parameters $\lambda_i$
coming from the definition of $\bar\rho$
(see e.g. lemma \ref{uhat}). Also, the dimension
of such a representation is induced by the map $\s_{gh}$ and
adds up to $N^{\Delta^{+}(g_n)}$. This gives rise to the following
theorem.

\begin{theorem} \label{smax}
Using inductively the representations of theorems \ref{t1}-\ref{t8}
to define representations of $\bar U_{\e}(A_1)$,
$\cdots,
\bar U_{\e}(g_{n-2})$,$\bar U_{\e}(g_{n-1})$ one obtains
representations of $U_{\e}(g_n)$ which have maximal dimensions and
number of free parameters for odd $N$, in all cases of semi simple
Lie algebras $g_n$.
\end{theorem}

\begin{remark}\rm
In the case of $U_{\e}(A_n)$ one obtains the maximal cyclic
representations of \cite{JI3}. In \cite{JI3} it is also proven,
that this representation is generically irreducible.
In the case of quantum $SO(5)$ irreducible
representations of maximal dimension
were obtained in \cite{AR3}. For the general $B_n, C_n$ and $D_n$
series  the representations of
maximal dimension of theorem \ref{smax} coincide
with those in \cite{SC}, for which the irreducibility for odd $N$ was
established in the simplest examples.
\end{remark}

\begin{remark}\rm
The construction of algebra homomorphisms and
representations for $U_q(g_n)$ was based on the action
of the generators $e_i,f_i,t_i^{\pm 1}$ on monomials of length
$\Delta^{+}(g_n)-\Delta^{+}(g_{n-1})$. Such monomials were defined
e.g. in example \ref{e2} and in (\ref{wais}). Adjoining these
monomials according to the inductive procedure of
theorem \ref{smax} gives monomials
$f_{i_1}^{(j_1)}\cdots f_{i_l}^{(j_l)}$ of length $l=
\Delta^{+}(g_n)$.
In all cases of quantum Lie algebras discussed above the ordering in these
monomials
corresponds to the ordering of simple reflexions in a longest element in the
Weyl group of $g_n$.
\end{remark}

\sect{{Conclusions}}
Starting from an arbitrary representations of $U_q(A_1)$ one can construct
representations for all higher rank semi simple Lie algebras by
``adding'' the additional generators which arise with the adding
of a new node to the Dynkin diagram.  The dependence of the
constructed representations of
$U_{\e}(g_n)$ on the algebra $U_{\e}(g_{n-1})$ results
in representations of quantum groups in dimensions greater than or equal to
$N^{\Delta^{+}(g_n)-\Delta^{+}(g_{n-1})}$. Only comparable very few
representations in such dimensions were previously known. Even
more representations, especially highest weight representations of
$U_{\e}(g_n)$ can be found by specializing some or all of
the free parameters. Although, the constructed
representations do not yet lead to representations of
quantum groups at $\e$ an $N$-th root of unity in all
possible dimensions  for irreducible representations, one
could hope that this problem might be settled in the future.

The minimal and maximal cyclic representations of $U_{\e}(A_n)$, which
are both special cases of the above described representations are
basic algebraic structures related to the generalizations of the chiral
Potts model in \cite{BA2,JI3,TA1}.
The next step is to find statistical models related to representations
of the other $\e$-deformed Lie algebras $g_n$ discussed above. In analogy
to the $U_q(sl_n)$
case one would expect that starting from the affine extension
of an $\e$-deformation of an arbitrary semi simple Lie algebra $g_n$ one
could find algebraic varieties which determine relations among
the spectral parameters to allow the existence and the construction
of the intertwining $R$-matrix of two representations.
This article is meant to be a small contribution to the research going
in this direction.

But applications of roots of unity representations are not
restricted to chiral Potts type models.
Many further applications
are found to lie  in the development of other
integrable models, conformal field theory (e.g. fusion rules),
 and areas in mathematics
such as the representation theory of
affine Lie algebras or the theory of semi-simple groups over fields
of positive characteristic.

{\bf Acknowledgement} I am greatly indebted to M. Jimbo and T. Miwa for
many discussions and their steady interest in this work,
and to the Research Institute of  Mathematical Sciences,
Kyoto University for its kind hospitality.

\sect{{ Appendix}}
In the case of applying the construction procedure of section
4 to derive algebra homomorphisms for $U_q(E_6), U_q(E_7)$
quantum Lie algebras only $U_q(A_2)$ type relations in
\ref{d1} were necessary.
For the construction of representations in the case of $E_8$ and
$F_4$ it is necessary to use two further  relations
in $U_q(E_8)$ and $U_q(F_4)$.
\begin{lemma} \label{l1}
Let $f_1,\dots, f_8$ be generators in $U_q(E_8)$. Then the generator
$f_1$ commutes with the monomial
$f_4^{(j_1)}f_5^{(j_2)}f_3^{(j_3)}f_4^{(j_4)}f_1^{(j_5)}
f_4^{(j_6)}f_5^{(j_7)}f_3^{(j_8)}f_4^{(j_9)}$ as following
$\gdef\[{\lbrack}$
$\gdef\]{\rbrack}$
\begin{eqnarray*}
&&f_1f_4^{(j_1)}f_5^{(j_2)}f_3^{(j_3)}f_4^{(j_4)}f_1^{(j_5)}
f_4^{(j_6)}f_5^{(j_7)}f_3^{(j_8)}f_4^{(j_9)}=\\&&
f_4^{(j_1-1)}f_5^{(j_2-1)}f_3^{(j_3-1)}f_4^{(j_4-1)}f_1^{(j_5)}
f_4^{(j_6+1)}f_5^{(j_7+1)}f_3^{(j_8+1)}f_4^{(j_9+1))}f_1+\\&&
f_4^{(j_1-1)}f_5^{(j_2-1)}f_3^{(j_3-1)}f_4^{(j_4-1)}f_1^{(j_5+1)}
f_4^{(j_6+1)}f_5^{(j_7+1)}f_3^{(j_8+1)}f_4^{(j_9+1)} \[-j_5+j_6+j_9+1\]+\\&&
f_4^{(j_1-1)}f_5^{(j_2-1)}f_3^{(j_3-1)}f_4^{(j_4)}f_1^{(j_5+1)}f_4^{(j_6+1)}
f_5^{(j_7+1)}f_3^{(j_8+1)}f_4^{(j_9)}\[ -j_4-j_6+j_8+j_7+1\]+\\&&
f_4^{(j_1-1)}f_5^{(j_2-1)}f_3^{(j_3-1)}f_4^{(j_4)}f_1^{(j_5+1)}
f_4^{(j_6)}f_5^{(j_7+1)}f_3^{(j_8+1)}f_4^{(j_9+1)}\[-j_4+j_9+1\]+\\&&
f_4^{(j_1-1)}f_5^{(j_2-1)}f_3^{(j_3)}f_4^{(j_4)}f_1^{(j_5+1)}f_4^{(j_6+1)}
f_5^{(j_7+1)}f_3^{(j_8)}f_4^{(j_9)}\[-j_3+j_7+1\]+\\&&
f_4^{(j_1-1)}f_5^{(j_2)}f_3^{(j_3-1)}f_4^{(j_4)}
f_1^{(j_5+1)}f_4^{(j_6+1)}f_5^{(j_7)}f_3^{(j_8+1)}f_4^{(j_9)}
\[-j_2+j_8+1\]+\\&&
f_4^{(j_1-1)}f_5^{(j_2)}f_3^{(j_3)}f_4^{(j_4)}f_1^{(j_5+1)}f_4^{(j_6+1)}
f_5^{(j_7)}f_3^{(j_8)}f_4^{(j_9)}\[-j_2-j_3+j_4+j_6+1\]+\\&&
f_4^{(j_1)}f_5^{(j_2)}f_3^{(j_3)}f_4^{(j_4-1)}f_1^{(j_5+1)}f_4^{(j_6+1)}
f_5^{(j_7)}f_3^{(j_8)}f_4^{(j_9)}\[-j_1+j_6+1\]+\\&&
f_4^{(j_1)}f_5^{(j_2)}f_3^{(j_3)}f_4^{(j_4)}f_1^{(j_5+1)}f_4^{(j_6)}
f_5^{(j_7)}f_3^{(j_8)}f_4^{(j_9)}\[-j_1-j_4+j_5+1\]
\end{eqnarray*}
\end{lemma}
The proof of this lemma is solely based on the $A_2$-type
Serre relations which
define the commutation relations among the $f_i$ generators.

\begin{lemma}\label{l2}
Let $f_1,\dots,f_4$ be generators of $U_q(F_4)$. Then the
action of $f_3$ on the monomial
$f_4^{(j_1)}f_3^{(j_2)}f_2^{(j_3)}f_1^{(j_4)}f_3^{(j_5)}
f_2^{(j_6)}f_3^{(j_7)}f_4^{(j_8)}f_3^{(j_9)}f_2^{(j_{10})}
f_3^{(j_{11})}$ is given by
\begin{eqnarray*}
&&f_3 f_4^{(j_1)}f_3^{(j_2)}f_2^{(j_3)}f_1^{(j_4)}f_3^{(j_5)}
f_2^{(j_6)}f_3^{(j_7)}f_4^{(j_8)}f_3^{(j_9)}f_2^{(j_{10})}
f_3^{(j_{11})}=\\&&
f_4^{(j_1-1)}f_3^{(j_2+1)}f_2^{(j_3)}f_1^{(j_4)}f_3^{(j_5-1)}
f_2^{(j_6-1)}f_3^{(j_7-1)}f_4^{(j_8)}f_3^{(j_9+1)}f_2^{(j_{10}+1)}
f_3^{(j_{11}+1)}f_4+\\&&
f_4^{(j_1-1)}f_3^{(j_2+1)}f_2^{(j_3)}f_1^{(j_4)}f_3^{(j_5-1)}
f_2^{(j_6-1)}f_3^{(j_7-1)}f_4^{(j_8+1)}f_3^{(j_9+1)}f_2^{(j_{10}+1)}
f_3^{(j_{11}+1)}\[j_{11}+j_9+1-j_8\]+\\&&
f_4^{(j_1-1)}f_3^{(j_2+1)}f_2^{(j_3)}f_1^{(j_4)}f_3^{(j_5-1)}f_2^{(j_6-1)}
f_3^{(j_7)}f_4^{(j_8+1)}f_3^{(j_9+1)}f_2^{(j_{10}+1)}f_3^{(j_{11})}
\[2j_{10}+1-j_7-j_9\]+\\&&
f_4^{(j_1-1)}f_3^{(j_2+1)}f_2^{(j_3)}f_1^{(j_4)}f_3^{(j_5-1)}f_2^{(j_6-1)}
f_3^{(j_7)}f_4^{(j_8+1)}f_3^{(j_9)}f_2^{(j_{10}+1)}f_3^{(j_{11}+1)}
\[j_{11}+1-j_7\]+\\&&
f_4^{(j_1-1)}f_3^{(j_2+1)}f_2^{(j_3)}f_1^{(j_4)}f_3^{(j_5-1)}f_2^{(j_6)}
f_3^{(j_7)}f_4^{(j_8+1)}f_3^{(j_9+1)}f_2^{(j_{10})}f_3^{(j_{11})}
\[-2j_6+j_7+j_9+1\]+\\&&
f_4^{(j_1-1)}f_3^{(j_2+1)}f_2^{(j_3)}f_1^{(j_4)}f_3^{(j_5)}f_2^{(j_6)}
f_3^{(j_7-1)}f_4^{(j_8+1)}f_3^{(j_9+1)}f_2^{(j_{10})}f_3^{(j_{11})}
\[-j_5+j_9+1\]+\\&&
f_4^{(j_1-1)}f_3^{(j_2+1)}f_2^{(j_3)}f_1^{(j_4)}f_3^{(j_5)}
f_2^{(j_6)}f_3^{(j_7)}f_4^{(j_8+1)}f_3^{(j_9)}f_2^{(j_{10})}f_3^{(j_{11})}
\[-j_5-j_7+j_8+1\]+\\&&
f_4^{(j_1)}f_3^{(j_2+1)}f_2^{(j_3)}f_1^{(j_4)}f_3^{(j_5)}f_2^{(j_6)}
f_3^{(j_7)}f_4^{(j_8)}f_3^{(j_9)}f_2^{(j_{10})}f_3^{(j_{11})}
\[-j_1+j_2+1\]
\end{eqnarray*}
\end{lemma}
The identity of this lemma was obtained using the commutation relations among
$f_i$
generators with terms $f_k^{(j)}$ in the cases of $U_q(A_2)$, $U_q(B_2)$
and $U_q(C_2)$.

With the help of the above lemmas it is possible
construct  the representation of $U_{\e}(E_8)$ and $U_q(F_4)$ in section 4.

\end{document}